\documentclass[useAMS]{mn2e}
\usepackage{times,epsfig,amsmath} 
\usepackage{graphicx}
\usepackage{epstopdf}
\usepackage{multirow}

\newcommand{\cxo}{{\it Chandra}}

\newcommand{\xmm}{{\it XMM-Newton}}

\title[The physics inside the X-ray/SZ scaling relations]
{The physics inside the scaling relations for X-ray galaxy clusters: \\ gas clumpiness, gas mass fraction and slope of the pressure profile}

\author[S. Ettori]
{S. Ettori$^{1,2}$ \\
\footnotesize 
 $^1$ INAF, Osservatorio Astronomico di Bologna, via Ranzani 1, I-40127 Bologna, Italy \\
 $^2$ INFN, Sezione di Bologna, viale Berti Pichat 6/2, I-40127 Bologna, Italy \\
 }

\date{Accepted 2014 October 29.  Received 2014 October 14; in original form 2014 May 14.}

\begin{document}
\maketitle 

\begin{abstract}
In galaxy clusters, the relations between observables in X-ray and millimeter wave bands and the total mass have normalizations, slopes and redshift evolutions that are simple to estimate in a self-similar scenario.
We study these scaling relations and show that they can be efficiently expressed, in a more coherent picture, by fixing the normalizations and slopes to the self-similar predictions, and advocating, as responsible of the observed deviations,  only three physical mass-dependent quantities:
the gas clumpiness $C$, the gas mass fraction $f_g$ and the logarithmic slope of the thermal pressure profile $\beta_P$. 
We use samples of the observed gas masses, temperature, luminosities, and Compton parameters in local clusters to constrain normalization and mass dependence of these 3 physical quantities, and measure: 
$C^{0.5} f_g = 0.110 (\pm 0.002 \pm 0.002) \left( E_z M / 5 \times 10^{14} M_{\odot} \right)^{0.198 (\pm 0.025 \pm 0.04)}$ and $\beta_P =  -d \ln P/d \ln r  = 3.14 (\pm 0.04 \pm 0.02)  \left( E_z M / 5 \times 10^{14} M_{\odot} \right)^{0.071 (\pm 0.012 \pm 0.004)}$, where both a statistical and systematic error (the latter mainly due to the cross-calibration uncertainties affecting the \cxo\ and \xmm\ results used in the present analysis) are quoted.
The degeneracy between $C$ and $f_g$ is broken by using the estimates of the Compton parameters.
Together with the self-similar predictions, these estimates on $C$, $f_g$ and $\beta_P$ define an inter-correlated internally-consistent set of scaling relations that reproduces the mass estimates with the lowest residuals. 
\end{abstract} 
 

\section{Introduction}

To use galaxy clusters as probes of the background Universe in which they form and evolve is essential to link some of their observed properties 
in the electromagnetic spectrum to their gravitational potential (see e.g. Allen, Evrard \& Mantz 2011, Kravtsov \& Borgani 2012).
Many proxies at different wavelengths, from radio to X-ray band, are nowadays available and robustly determined.
Some attempts are already started to combine few of these proxies to improve the constraints on the inferred mass
(e.g. Stanek et al. 2010, Okabe et al. 2010, Ettori et al. 2012, Ettori 2013, Maughan 2014, Rozo et al. 2014, Evrard et al. 2014).

In this work, we focus on the Intra-Cluster Medium (ICM), the hot fully-ionized optically-thin plasma that collapses into the cluster gravitational potential. 
The physical processes occurring in the ICM can be mapped both with the X-rays, produced via bremsstrahlung radiation (e.g. B\"ohringer \& Werner 2010), 
and through the Sunyaev-Zeldovich (hereafter SZ) effect, that traces the Compton scattering 
of the photons of the Cosmic Microwave Background on the electrons of the same plasma (Sunyaev \& Zeldovich 1980).

In particular, we consider the scaling relations between cluster masses and the X-ray/SZ observables (see Giodini et al. 2013 for a recent review on this topic). 
We obtain, first, the analytic expressions that relate gas mass, temperature, luminosity and Compton parameter to the total mass and,
then, we show that these relations, with the normalizations and slopes fixed to the analytic values, can be used more efficiently to estimate the total mass, once
a set of 3 physically-motivated quantities are defined also in their mass dependence.

The paper is organized as follows. In Section~2, we introduce the scaling relations considered for our analysis, providing a numerical value for the normalization that depends just on three unknown quantities, i.e. the average gas clumpiness, the cluster gas mass fraction and the slope of the gas pressure profile.
In Section~3, we describe how we can calibrate the investigated scaling relations by using the largest sample available of hydrostatic mass measurements.
In Section~4, we summarize our main findings.
Hereafter, all the physical quantities considered refer to the cosmological parameters $H_0=70$ km s$^{-1}$ Mpc$^{-1}$ and $\Omega_{\rm m} = 1 - \Omega_{\Lambda}=0.3$, unless stated otherwise. 

\section{The X-ray and SZ scaling laws for the total mass}

For a galaxy cluster in hydrostatic equilibrium, the radial profile of the total mass is described by the equation (e.g. Ettori et al. 2013)
\begin{equation}
M(<R) \equiv M = -\frac{R \; T(R)}{\mu m_p G} \frac{d \ln P}{d \ln r} =  \frac{R \; T \; f_T \; \beta_P}{\mu m_p G},
\end{equation}
where $\beta_P = -d \ln P/d \ln r >0$ is the opposite of the logarithmic slope of the gas pressure profile, and
$f_T = T(R) / T$ is defined as the ratio between the 3D value of the gas temperature at the radius $R$ and the mean spectroscopic estimate $T$,
that will appear in the scaling relations.

\begin{table*}
\caption{Properties of the multi-wavelength samples considered in the present analysis. In the columns $M$ and $X$, 
the median value, the range covered (in parentheses) and the relative error of the mass and the investigate observable, respectively,
are quoted. The units for the observables $X$ are: $10^{13} M_{\odot}$ for $M_g$; keV for $T$; $10^{44}$ erg s$^{1}$ for $L$; 
$D_A^2 10^4$ Mpc$^{-2}$ for $Y_{SZ}$.
} \begin{tabular}{ccccc} \hline
 Sample & $N$ & $z$ & $M$ & $X$  \\ 
 & & & $10^{14} M_{\odot}$ &  \\ \hline
 All $M$ & 213 & $0.226 \; (0.012-1.390)$ & $3.51 \; (0.15-22.80); \; 0.19$ & $-$ \\
 $M-M_g$  & 109 & $0.141 \; (0.012-0.550)$ & $3.15 \; (0.15-14.52); \; 0.19$ & $3.91 \; (0.08-26.7); \; 0.09$ \\
 $M-T$  & 213 & $0.226 \; (0.012-1.390)$ & $3.51 \; (0.15-22.80); \; 0.19$ & $5.30 \; (0.81-12.5); \; 0.07$ \\
 $M-L$  & 199 & $0.231 \; (0.012-1.390)$ & $3.52 \; (0.15-22.80); \; 0.20$ & $6.80 \; (0.02-118.2); \; 0.04$ \\
 $M-Y_{\rm SZ}$  & 94 & $0.176 \; (0.048-0.548)$ & $5.99 \; (0.98-14.52); \; 0.19$ & $0.66 \; (0.06-3.8); \; 0.14$ \\
\hline \end{tabular}

\label{tab:prop}
\end{table*}

Studies of the properties of the self-similar scaling scenario have shown to be more convenient to refer to cluster's regions defined with respect a fixed overdensity when halo with different masses and redshifts are considered (e.g. B\"ohringer et al. 2012).
In our analysis, we consider physical quantities estimated within a radius $R_{\Delta}$, that defines a spherical region where the mean mass overdensity $\Delta$ is evaluated with respect to the critical density of the Universe at the cluster's redshift $z$, $\rho_{c,z} = 3 H_z^2 / (8 \pi G)$: $\Delta = 3 M / (4 \pi \rho_{c,z} R_{\Delta}^3) = 2 G M / (H_z^2 R_{\Delta}^3)$, where the Hubble constant $H_z = H_0 E_z$ includes the factor describing its cosmic evolution $E_z = \left[\Omega_{\rm m} (1+z)^3 + 1 - \Omega_{\rm m} \right]^{1/2}$ for a flat cosmology with matter density parameter $\Omega_{\rm m}$. 

By assuming 
\begin{enumerate}
\item a gas mass fraction $f_g = M_g / M$;
\item that the X-ray emission is mostly due to bremsstrahlung processes so that the bolometric luminosity $L \equiv L_{\rm bol} = \int{ n_e n_p \Lambda(T) dV} = f_L M_g^2 c_f / (\mu_e^2 m_{\rm amu}^2 V)$, where: $V = 4/3 \pi R_{\Delta}^3$ is the cluster volume; $f_L = \int{n_g^2 dV} / (\int{n_g dV})^2 \, V$ is the correction needed to consider the gas mass ($\int{n_g dV}$) instead of the emission integral ($\int{n_g^2 dV}$) for the scaling purpose and is equal to 1.80 for a gas density distribution described by a $\beta-$model with $\beta=0.65$ and $R_{500} = 5 \times$ the core radius, that are the median values of the estimated parameters in the sample of the brightest 45 nearby galaxy clusters in Mohr et al. (1999)\footnote{Using the extremes of the inter-quartile ranges of the estimated values of $\beta$ and $R_{500}$ in the Mohr et al. sample, we estimate the variations on $f_L$ between $-15$ and $+29$ per cent, and, through the dependence to the power of $-3/4$, on the quoted normalization of the $M-L$ relation between $-17$ and $+13$ per cent.};  the cooling function $c_f$ is equal to $c_{f,0} \times T_{\rm keV}^{0.5}$ erg s$^{-1}$ cm$^{3}$, with $c_{f,0} = 0.85 \times 10^{-23} c_{pe}$ (this value is completely consistent with, e.g., Sutherland \& Dopita 1993 as tabulated in Tozzi \& Norman 2001)\footnote{The normalization of the cooling function is estimated by fitting a function $c T^{0.5}$ to the values of the cooling function evaluated with the thermal model {\tt apec} in XSPEC (Arnaud 1996) where a metallicity of 0.3 times the solar abundance as tabulated in Anders \& Grevesse (1989) and a set of temperature between 2 and 12 keV are considered; adopting a metallicity of 0.1 decreases the normalization of $c_f$ by 5 per cent; the difference is --6 per cent when a metallicity of 0.3 and the more recent table of solar abundance from Asplund et al. (2009) are considered.}; a conversion factor from protons to electrons $c_{pe} = 1.1995$, an electronic weight $\mu_e=1.1738$ and an atomic mass $m_{\rm amu} = 1.66 \times 10^{-24}$ g are used;
\item that the millimeter wave emission is due to the SZ effect which is proportional to the integrated pressure of the X-ray emitting plasma along the line-of-sight and is described from the integrated Compton parameter $Y_{SZ} D_A^2 = (\sigma_T/ m_e c^2) \int P dV$, where $D_A$ is the angular distance to the cluster, $\sigma_T = 8 \pi/3 ( e^2 / m_e c^2 )^2 = 6.65 \times 10^{-25}$ cm$^2$ is the Thompson cross section, $m_e$ and $e$ are the electron rest mass and charge, respectively, $c$ is the speed of light, and $P = n_e T$ is the electron pressure profile;
\end{enumerate}
we can write the following scaling laws with their calculated normalization
\begin{align}
\frac{F_z M}{5 \times 10^{14} M_{\odot}} & = 1.0 \left( \frac{C^{0.5} \, f_g}{0.1} \right)^{-1} \frac{F_z M_g}{5 \times 10^{13} M_{\odot}}  \nonumber \\
 & \hspace*{-0.8cm}  = 0.832 \left( \frac{\beta_P}{3} \right)^{3/2} \left( \frac{kT}{5 keV} \right)^{3/2} \nonumber \\
 & \hspace*{-0.8cm}  = 0.962 \left( \frac{\beta_P}{3} \right)^{3/8} \left( \frac{C^{0.5} \, f_g}{0.1} \right)^{-3/2}  \left( \frac{F_z^{-1} L}{5\times10^{44} {\rm erg/s}} \right)^{3/4} \nonumber \\
 & \hspace*{-0.8cm}  = 1.748 \left( \frac{\beta_P}{3} \right)^{3/5} \left( \frac{f_g}{0.1} \right)^{-3/5}  \left( \frac{F_z Y_{SZ} D_A^2}{10^{-4} {\rm Mpc}^2} \right)^{3/5}.
\label{eq:sl}
\end{align}
Here, we define: $F_z = E_z \times (\Delta/500)^{0.5}$ (see e.g. Ettori et al. 2004); the clumpiness in the gas density $C = <n_g^2> / <n_g>^2$ that affects the measurement of the gas density as obtained from the deprojection of the X-ray data produced from free-free emission, but not from SZ signal due to inverse Compton (see e.g. Roncarelli et al. 2013, Eckert et al. 2013a, b); the mean atomic weight $\mu=0.61$. 
As reference values, we adopt: an overdensity of $\Delta=500$ (and therefore $F_z = E_z$), for which $f_T \approx 0.67$ (e.g. Vikhlinin et al. 2006, Baldi et al. 2012); a gas fraction of 0.1 (see e.g. Ettori et al. 2009, Mantz et al. 2014); a logarithmic slope of the gas pressure profile at $R_{500}$ of $-3$, which is consistent with the values in the range $(-3.2, -2.8)$ of the profiles adopted in Arnaud et al. (2010) and in the papers of the Planck  collaboration (2013).
We refer to the appendix for further details on how the normalizations are estimated (Sect.~\ref{sect:nor}) and to the extension of the $M-L$ relation to no-bolometric energy bands (Sect.~\ref{sect:ml_eband}).

Following Ettori (2013  --hereafter E13), where a generalised form for the scaling laws has been presented, a concise form of all the set of the above equations can be written as
\begin{equation}
F_z M  \sim \beta_P^{\theta} f_g^{-\phi} \; (F_z^{-1} L)^{\alpha} (F_z M_g)^{\beta} T^{\gamma} 
\label{eq:gsl}
\end{equation}
where the relations
\begin{align}
4 \alpha & +3 \beta +2 \gamma=3 \nonumber \\
\theta & = \alpha/2 +\gamma \nonumber \\
\phi & = 2 \alpha + \beta
\label{eq:gsl_rel}
\end{align}
among the exponents hold in a self-similar scenario (e.g., the $M-T$ relation is recovered by imposing the absence of any dependence on $M_g$ and $L$, i.e. $\alpha=\beta=0$; then, $\gamma=3/2$, $\theta=\gamma=3/2$ and $\phi=0$), and $Y_{SZ}$ is here represented as the product of gas mass and temperature.

\begin{figure*}
\hbox{
  \includegraphics[width=0.5\textwidth, keepaspectratio]{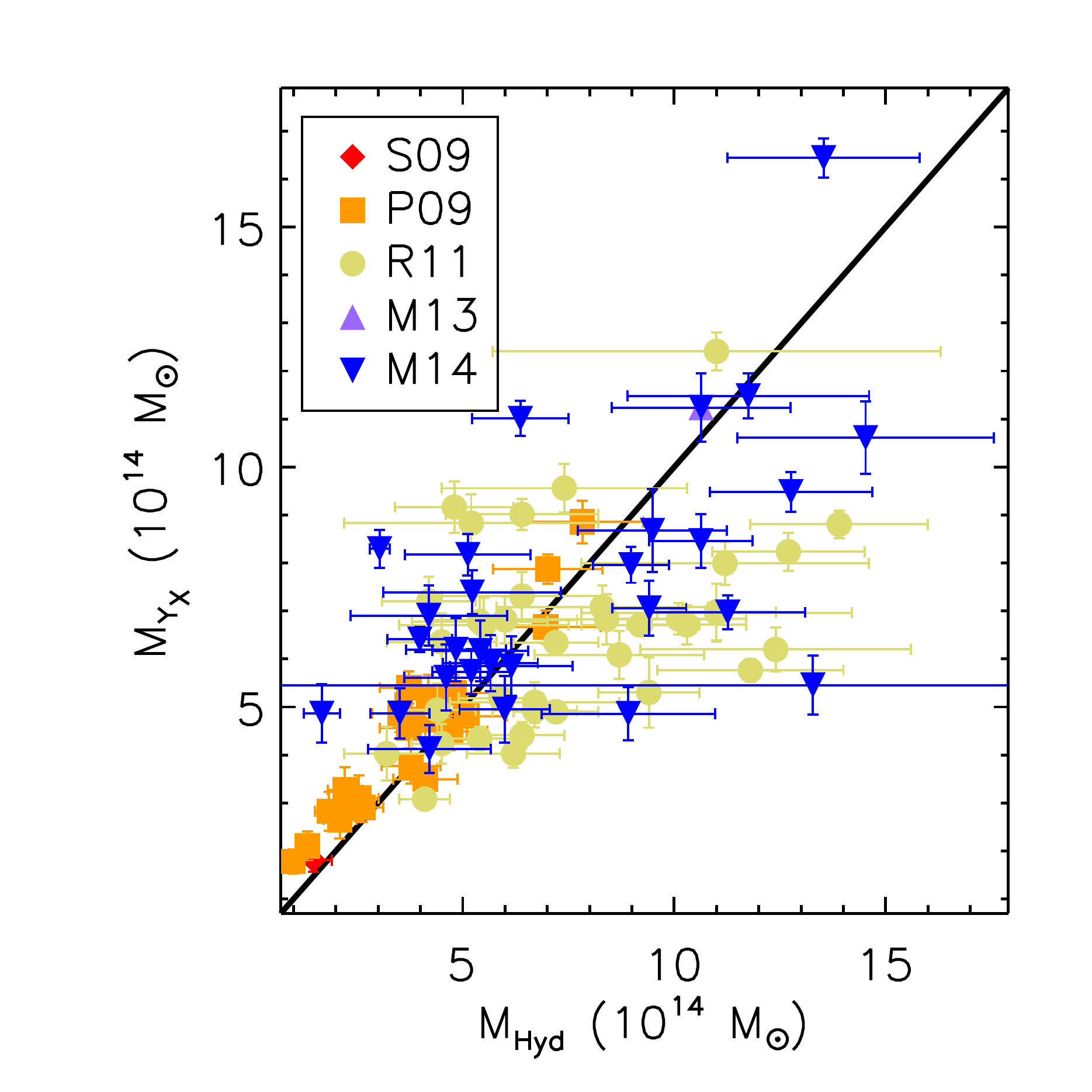}
  \includegraphics[width=0.5\textwidth, keepaspectratio]{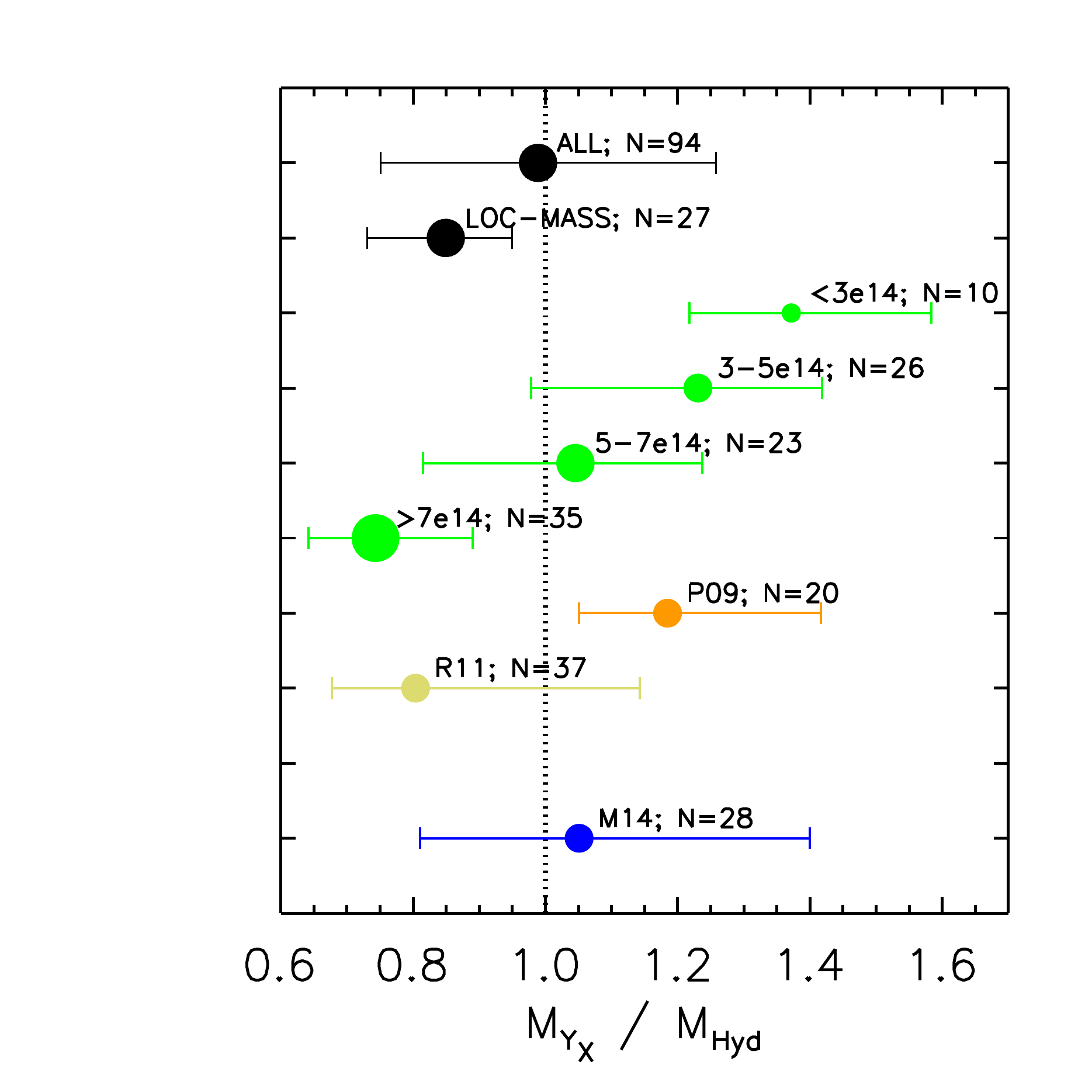}
}  \caption{(Left) Comparison between our collection of hydrostatic masses (S09=Sun et al. 2009; P09=Pratt et al. 2009; R11=Reichert et al. 2011; M13=Mahdavi et al. 2013; M14=Maughan 2014) and $M_{Y_X}$ as estimated in Planck  collaboration (2014).
(Right) Median values (with the lower and upper quartile) of the ratios $M / M_{Y_X}$ for the whole sample (black), in 4 different mass bins (green) and for the different considered datasets.
} \label{fig:planck_m}
\end{figure*}

\section{The calibration of the scaling relations}
 \label{sect:samples}

To check the consistency between the normalizations in equation~\ref{eq:sl} and the observed distributions, we consider X-ray mass estimates obtained through the application of the equation of the hydrostatic equilibrium under the assumptions that any gas velocity is zero and that the ICM is distributed in a spherically symmetric way into the cluster gravitational potential (see, e.g., Ettori et al. 2013). 
We start with the sample described in Ettori (2013; 120 entries, 113 of which are unique hydrostatic mass measurements collected from the public catalogs in Sun et al. 2009, Pratt et al. 2009\footnote{The masses derived in Pratt et al. (2009) for the objects in the REXCESS  sample are not obtained from the equation of the hydrostatic equilibrium, but are estimated from the $Y_X (= M_g T)-M$ relation as calibrated in Arnaud et al. (2007). We consider them in our sample for the wealth of information associated to the REXCESS catalog.}, Mahdavi et al. 2013, Maughan 2014) and add the 110 (out of 232) objects present in the Reichert et al. (2011) sample and not considered in E13, for a total number of 213 galaxy clusters with reliable hydrostatic masses estimated at $\Delta=500$.
Using this sample, we investigate, as described below, the normalization and slope of the $M-T$ relation (213 objects), $M-L$ relation (199 objects), $M-M_{\rm gas}$ relation (113 objects).
Then, we consider the Planck catalog (file {\it  COM\_PCCS\_SZ-validation\_R1.13.fits} available at {\tt http://www.sciops.esa.int/index.php?page =Planck\_Legacy\_Archive\&project=planck}; see Planck  collaboration 2014) with 1227 entries, 455 of which with estimated redshift and $Y_{SZ} \equiv Y_{500, PSX} >0$.
We obtain that 94 are the systems in common between the 213 galaxy cluster with hydrostatic masses and the 455 Planck clusters.
In the published catalog, also estimates of the mass, $M_{Y_X}$, obtained through the $Y_X = M_g T$ parameter (e.g. Kravtsov et al. 2006, Arnaud et al. 2010) are provided.
For the 94 objects in common, we calculate the ratio between the collected values of the hydrostatic mass and $M_{Y_X}$. 
We obtain an overall perfect agreement (median value: 0.99). On the other hand, we also notice a clear bias depending on the total hydrostatic mass, with systems at lower ($<3 \times 10^{14} M_{\odot}$) and higher ($>7 \times 10^{14} M_{\odot}$) masses showing the highest deviations (median values of 1.37 and 0.74, respectively; see Fig.~\ref{fig:planck_m}), indicating that the collected hydrostatic estimates over (under) predict the high (low) values of $M_{Y_X}$.

The main properties of the sample here analyzed are listed in Table~\ref{tab:prop}.

We fit these quantities using the linear function 
\begin{equation}
\mathcal{Y} = n \; +a \mathcal{X}  
 \label{eq:fit}
\end{equation}
and minimizing the merit function
\begin{align}
 \chi^2 = & \sum_{i=1}^N \frac{(\mathcal{Y}_i \; -n \; -a \mathcal{X}_i)^2}{ \epsilon_i^2}  \nonumber \\
 \epsilon_i^2 = & \epsilon_{\mathcal{Y}, i}^2 \; +a^2 \epsilon_{\mathcal{X}, i}^2 \; -2 \, a \, \rho \, \epsilon_{\mathcal{Y}, i} \, \epsilon_{\mathcal{X}, i},
 \label{eq:chi2}
\end{align}
where $\mathcal{Y} = \log\left( \frac{F_z \; M}{5 \times 10^{14} M_{\odot}} \right), \mathcal{X} = \log(X)$ and  
$X$ is equal to $\frac{F_z \; M_{g}}{5 \times 10^{13} M_{\odot}}$, $\frac{T}{5 {\rm keV}}$, $ \frac{F_z^{-1} \;  L_{\rm bol}}{5\times10^{44} {\rm erg/s}}$,
$\frac{F_z Y_{SZ} D_A^2}{10^{-4} {\rm Mpc}^2}$; 
``$\log$'' indicates the base-10 logarithm;
the associated errors $\epsilon_{\mathcal{Y}}$ and $\epsilon_{\mathcal{X}}$ are obtained through the propagation of the measured uncertainties;
$N$ is the number of data points and $D=N-p$ are the degrees of freedom given a number $p$ of fitted parameters (either 2 --slope and normalization-- or the normalization only); $\rho$ is the Pearson's correlation coefficient among the variables $\mathcal{Y}$ and $\mathcal{X}$.
An intrinsic scatter is estimated by adding it in quadrature to $\epsilon_i$ and re-iterating the fitting procedure until a reduced $\chi^2$ of 1 is obtained.
The relative error on it is obtained as discussed in E13.
The fit is performed using the {\tt IDL} routine {\it MPFIT} (Markwardt 2008).



Although we provide all the calculations needed to investigate the evolution with redshift of the scaling relations, we prefer not to study it in the present work because of the heterogenous origin of the considered dataset that, without a proper weight provided from a redshift-dependent selection function, could affect any conclusion on the redshift evolution.

\begin{table*}
\caption{Best-fit results for the scaling relations investigated. $N$ is the number of fitted data; $\rho$ is the Pearson' correlation coefficient; $n$ and $a$ refer to equation~\ref{eq:fit}; $\chi^2_r$ is the reduced $\chi^2$; $\sigma_i$ is the intrinsic scatter in $\log M$ at given observable $ \mathcal{X}$; $m$ and $\mathcal{N}^c$ are described in equation~\ref{eq:cor} and are obtained from the fits with normalisation $n$ and slope $a$ as free parameters. 
For each scaling relation, we provide the best-fit results obtained with: 
{\it (1)} eq.~\ref{eq:fit} with the slope fixed to the self-similar value and the sub-sample of ``local'' (i.e. $z<0.15$) and ``massive'' (i.e. $M_{500}>3 \times 10^{14} M_{\odot}$) objects; 
{\it (2)}  as for {\it (1)}, but leaving the slope free to vary; 
{\it (3)} eq.~\ref{eq:fit_m}; 
{\it (4-6)} as for {\it (1-3)}, but for the subsample of the all ``local'' clusters.
In particular, the 3rd row of the ``local'' samples (i.e. fit {\it (3)} e {\it (6)}) refers to the best-fit results for a fixed slope and including the mass-dependence described in equation~\ref{eq:fit_m}. 
Finally, the fit {\it (7)} is obtained by equation~\ref{eq:fit_m} with the normalization fixed after the calibration of the physical quantities described in eq.~\ref{eq:par_m} (see the instructions at the end of Sect.~\ref{sect:bestfit}). 
In Fig.~\ref{fig:fit}, we show the samples and the best-fit lines that represent the results for the fits {\it (4-7)}.}
 \begin{tabular}{c@{\hspace{.6em}} c@{\hspace{.6em}} c@{\hspace{.6em}} c@{\hspace{.5em}} c@{\hspace{.5em}} c@{\hspace{.8em}}  c@{\hspace{.5em}} c@{\hspace{.5em}} c@{\hspace{.5em}}} \hline
 Sample & $N$ & $\rho$ & $10^n = \mathcal{N}_{\rm obs}$ & $a$ & $\chi^2_r$ & $\sigma_i$ & $m$ & $\mathcal{N}^c$ \\ \hline
\multicolumn{9}{c}{  $M-M_g$  } \\
 (1) {\bf local massive} & 16 & 0.94 & $0.883\pm0.012$ & $1.000$ & 6.9 & $0.053\pm0.013$ & $-$ & $-$ \\
 (2) &  & & $0.923\pm0.033$ & $0.751\pm0.067$ & 2.3 & $0.028\pm0.023$ & $-0.332\pm0.119$ & $0.899\pm0.112$ \\
 (3) (eq.~\ref{eq:fit_m}) &  & & $0.912\pm0.017$ & $1.000$ & 1.4 & $0.019\pm0.016$ & $-$ & $-$ \\
 (4) {\bf local all} & 59 & 0.97 & $1.065\pm0.005$ & $1.000$ & 44.4 & $0.096\pm0.010$ & $-$ & $-$ \\
 (5) &  & & $0.848\pm0.025$ & $0.835\pm0.017$ & 7.8 & $0.048\pm0.007$ & $-0.198\pm0.025$ & $0.821\pm0.031$ \\
 (6) (eq.~\ref{eq:fit_m}) &  & & $0.912\pm0.011$ & $1.000$ & 1.0 & $0.000\pm0.003$ & $-$ & $-$ \\
 (7) &  & & $0.912$ & $1.000$ & 1.0 & $0.000\pm0.004$ & $-$ & $-$ \\
 &  & & \\
\multicolumn{9}{c}{  $M-T$  } \\
 (1) {\bf local massive} & 29 & 0.90 & $0.927\pm0.017$ & $1.500$ & 1.7 & $0.043\pm0.016$ & $-$ & $-$ \\
 (2) &  & & $0.947\pm0.034$ & $1.390\pm0.119$ & 1.6 & $0.044\pm0.018$ & $-0.053\pm0.062$ & $0.943\pm0.094$ \\
 (3) (eq.~\ref{eq:fit_m}) &  & & $0.890\pm0.017$ & $1.500$ & 1.9 & $0.042\pm0.014$ & $-$ & $-$ \\
 (4) {\bf local all} & 73 & 0.97 & $0.780\pm0.008$ & $1.500$ & 4.9 & $0.073\pm0.009$ & $-$ & $-$ \\
 (5) &  & & $0.873\pm0.021$ & $1.679\pm0.033$ & 3.6 & $0.055\pm0.008$ & $0.071\pm0.012$ & $0.886\pm0.027$ \\
 (6) (eq.~\ref{eq:fit_m}) &  & & $0.881\pm0.014$ & $1.500$ & 1.0 & $0.007\pm0.011$ & $-$ & $-$ \\
 (7) &  & & $0.890$ & $1.500$ & 1.1 & $0.009\pm0.011$ & $-$ & $-$ \\
 &  & & \\
\multicolumn{9}{c}{  $M-L$  } \\
 (1) {\bf local massive} & 22 & 0.87 & $0.830\pm0.022$ & $0.750$ & 2.2 & $0.060\pm0.019$ & $-$ & $-$ \\
 (2) &  & & $0.855\pm0.054$ & $0.679\pm0.088$ & 2.2 & $0.061\pm0.021$ & $-0.138\pm0.190$ & $0.841\pm0.131$ \\
 (3) (eq.~\ref{eq:fit_m}) &  & & $0.909\pm0.035$ & $0.750$ & 1.6 & $0.070\pm0.031$ & $-$ & $-$ \\
 (4) {\bf local all} & 60 & 0.90 & $1.118\pm0.017$ & $0.750$ & 11.1 & $0.158\pm0.017$ & $-$ & $-$ \\
 (5) &  & & $0.821\pm0.046$ & $0.609\pm0.028$ & 12.6 & $0.103\pm0.017$ & $-0.309\pm0.076$ & $0.784\pm0.063$ \\
 (6) (eq.~\ref{eq:fit_m}) &  & & $0.837\pm0.023$ & $0.750$ & 2.5 & $0.122\pm0.020$ & $-$ & $-$ \\
 (7) &  & & $0.852$ & $0.750$ & 2.6 & $0.124\pm0.020$ & $-$ & $-$ \\
 &  & & \\
\multicolumn{9}{c}{  $M-Y_{\rm SZ}$  } \\
 (1) {\bf local massive} & 27 & 0.79 & $1.960\pm0.038$ & $0.600$ & 4.0 & $0.089\pm0.017$ & $-$ & $-$ \\
 (2) &  & & $1.923\pm0.169$ & $0.578\pm0.093$ & 4.1 & $0.091\pm0.018$ & $-0.063\pm0.277$ & $1.971\pm0.371$ \\
 (3) (eq.~\ref{eq:fit_m}) &  & & $2.113\pm0.052$ & $0.600$ & 2.5 & $0.080\pm0.021$ & $-$ & $-$ \\
 (4) {\bf local all} & 36 & 0.90 & $1.753\pm0.024$ & $0.600$ & 9.6 & $0.110\pm0.016$ & $-$ & $-$ \\
 (5) &  & & $2.165\pm0.369$ & $0.774\pm0.127$ & 9.2 & $0.098\pm0.032$ & $0.374\pm0.212$ & $1.820\pm0.387$ \\
 (6) (eq.~\ref{eq:fit_m}) &  & & $2.083\pm0.044$ & $0.600$ & 2.9 & $0.101\pm0.024$ & $-$ & $-$ \\
 (7) &  & & $2.113$ & $0.600$ & 2.9 & $0.103\pm0.024$ & $-$ & $-$ \\
 &  & & \\
\hline \end{tabular}

\label{tab:res}
\end{table*}

\subsection{A mass dependent deviation from self-similarity}
\label{sect:mass}

Since the first evidences of the deviations of the observed slopes of the X-ray scaling laws from the self-similar expectations,
it has been suggested that a possible solution to reconcile the predicted and observed values can be obtained
by assuming that at least one of the physical quantities (like, e.g. the gas mass fraction) appearing in the derivation of the scaling law has a not-negligible mass dependence (see e.g. Arnaud \& Evrard 1999, Pratt et al. 2009).
In this section, we investigate how we can constrain the mass dependence of the set of the physical quantities we need for a complete description of the 
scaling relations, by imposing that this mass dependence is fully responsible for any observed deviation from the self-similar prediction.
 
In general, we can write the scaling relations here investigated between the mass $M$ and an observable $X$ 
as $M = \mathcal{N}^c \; X^a \; E_z^b$.
Note that,  for sake of completeness, we are considering a normalization with its own exponent (to treat the case of, e.g., $\beta_P$ and $f_g$
in equation~\ref{eq:sl}) and a term for the redshift evolution proportional to $E_z$.
In the case that the normalization $\mathcal{N}$ depends upon the mass $M$, $\mathcal{N} = \mathcal{N}_0 M^m$, 
the scaling relations are then modified accordingly:
\begin{align}
M & = \mathcal{N}_0^{c/(1- c \, m)} \; X^{a/(1-c \, m)} \; E_z^{b/(1-c \,m)} = \mathcal{N}_{\rm obs} \; X^{a_{\rm obs}} \; E_z^{b_{\rm obs}},
\label{eq:xm}
\end{align}
where the subscript {\it obs} in the last member refers to the values measured by a best-fit procedure leaving normalization ($\mathcal{N}_{\rm obs}$), 
slope ($a_{\rm obs}$) and redshift evolution ($b_{\rm obs}$) free to vary.
By imposing $c=a=a_{\rm exp}$, where $a_{\rm exp}$ is the nominal exponent predicted in the self-similar scenario, and equating the 2nd and 3rd member of equation~\ref{eq:xm}, we obtain that the ``intrinsic", mass-corrected scaling relations can be recovered by estimating:
\begin{align}
m & = 1/a_{\rm exp} \; - \; 1/a_{\rm obs}  \nonumber \\
\mathcal{N}^c & =  \mathcal{N}_{\rm obs}^{1- c \, m}  = \mathcal{N}_{\rm obs}^{a_{\rm exp}/a_{\rm obs}} \nonumber \\
b & = b_{\rm obs} (1-c \, m) = b_{\rm obs} \, a_{\rm exp} / a_{\rm obs}.
\label{eq:cor}
\end{align}

Following the expression of equation~\ref{eq:sl}, where all the scaling relations can be written as function of $(E_z M)$, we assume a similar dependence on the mass of the 3 unknown parameters (i.e. the logarithmic slope of the pressure $\beta_P$, the gas mass fraction $f_g$ and the gas clumpiness $C$):
\begin{align}
\beta_P & = \beta_{P, 0} \; (E_z M)^{m_1} \nonumber  \\
f_g & = f_{g, 0} \; (E_z M)^{m_2} \nonumber  \\
C & = C_0 \; (E_z M)^{m_3}.
\label{eq:par_m}
\end{align}

Using equation~\ref{eq:par_m}, together with equations~\ref{eq:gsl} and \ref{eq:xm}, we can then write a general expression in the form
\begin{align}
\left( E_z M \right)^{1 -\theta \; m_1 + \phi \; m_2 +\phi \; m_3/2} \sim & \beta_{P, 0}^{\theta} C_0^{-\phi/2} f_{g, 0}^{-\phi} \nonumber \\
  & (E_z^{-1} L)^{\alpha} (E_z M_g)^{\beta} T^{\gamma}, 
\label{eq:sl_m}
\end{align}
that can be resolved in each of the scaling laws considered here (see eq.~\ref{eq:sl} and \ref{eq:gsl_rel}) as
\begin{align}
E_z M  & \sim \left( C_0^{0.5} f_{g, 0} \right)^{-1/(1-m)} \left(E_z M_g \right)^{1/(1-m)} \; ; \, m= -m_2 -m_3/2 \nonumber \\
 & \sim \beta_{P, 0}^{3/2 / (1-3/2 \; m)}  \left( kT \right)^{3/2 / (1-3/2 \; m)} \; ; \,  m=m_1 \nonumber \\
 & \sim \beta_{P, 0}^{3/8 / (1-3/4 \; m)}  f_{g, 0}^{-3/2 / (1-3/4 \; m)}   \nonumber \\
 & \; \left( E_z^{-1} L \right)^{3/4 / (1 -3/4 \; m)} \; ; \, m = m_1/2 -2 m_2 -m_3 \nonumber \\
 & \sim \beta_{P, 0}^{3/5 / (1-3/5 \; m)}  f_{g, 0}^{-3/5 / (1-3/5 \; m)}   \nonumber \\
  & \; \left(  E_z Y_{SZ} D_A^2 \right)^{3/5 / (1 -3/5 \; m)} \; ; \,  m = m_1 -m_2.
\label{eq:sl_msingle}
\end{align}
Here, the symbol ``$\sim$'' is used to replace all the factors and pivot values shown in eq.~\ref{eq:sl}.
These equations show explicitly the quantities that can be constrained by fitting a linear function, with normalization and slope as free parameters, to the logarithmic values of the mass and of the observables. For example, by fitting the $M-T$ relation, one can directly estimate $m=m_1$ from the best-fit value of the slope and $\beta_{P, 0}$ from the best-fit value of the normalization.
In the following subsection, we show how we can constrain the parameters of our interest, defined in eq.~\ref{eq:par_m}, by combining the results obtained from the linear fit of the scaling relations and quoted in Table~\ref{tab:res}.

Once the dependence on the mass is assessed, we can fit the scaling relation by fixing the expected slope $a_{\rm exp}$ and propagating the correction to the total mass:
\begin{equation}
(1 -\theta \; m_1 + \phi \; m_2 +\phi \; m_3/2) \log (E_z M) = \bar{n} +a_{\rm exp} \log(X).
\label{eq:fit_m}
\end{equation}
The normalization $\bar{n}$ is the only free parameter and is used to calibrate finally the gas mass fraction, $f_{g, 0}$, the gas clumpiness, $C_0$, and the logarithmic slope of the gas pressure, $\beta_{P, 0}$. 

\begin{figure*}
\begin{center}
\hbox{
  \includegraphics[width=0.5\textwidth, keepaspectratio]{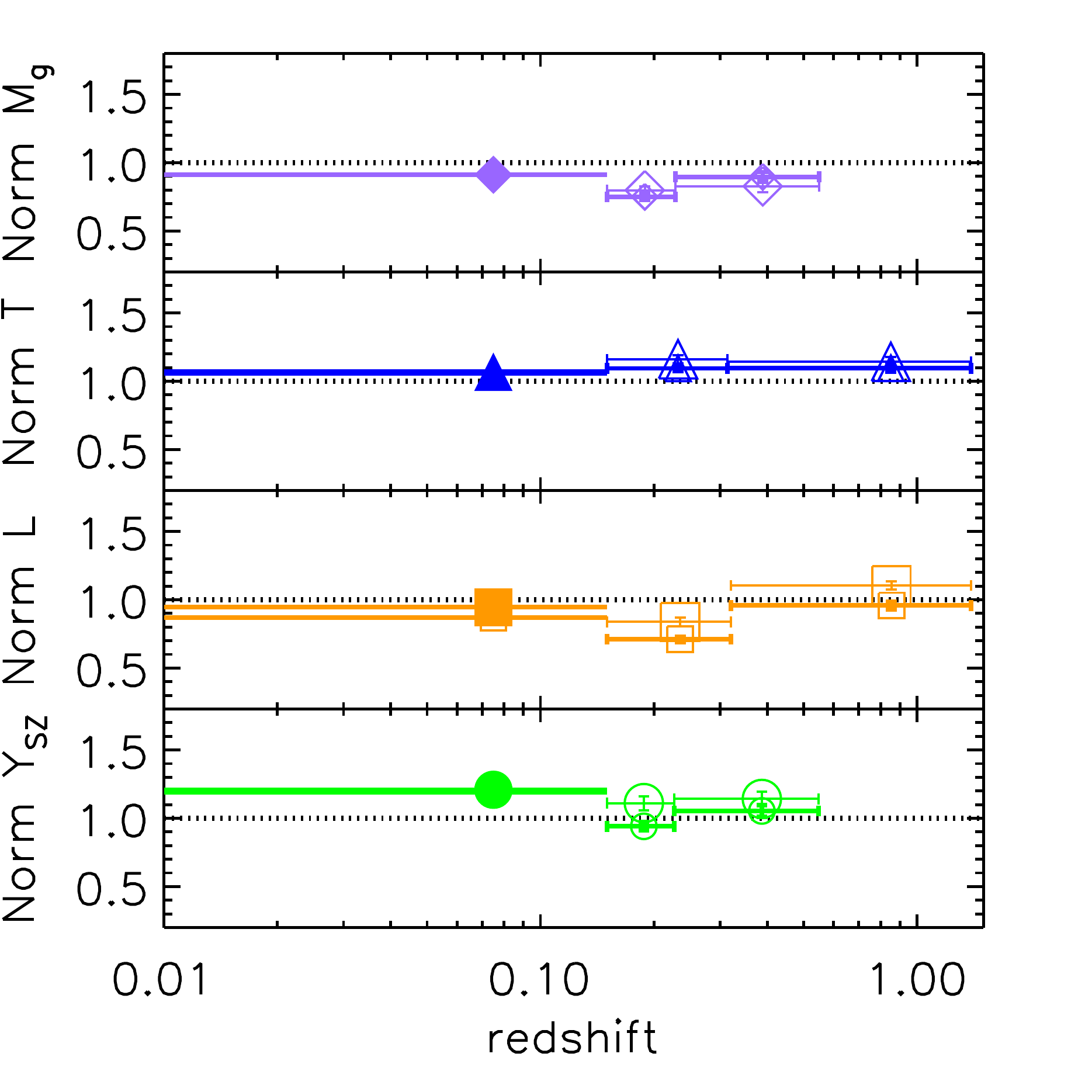}
  \includegraphics[width=0.5\textwidth, keepaspectratio]{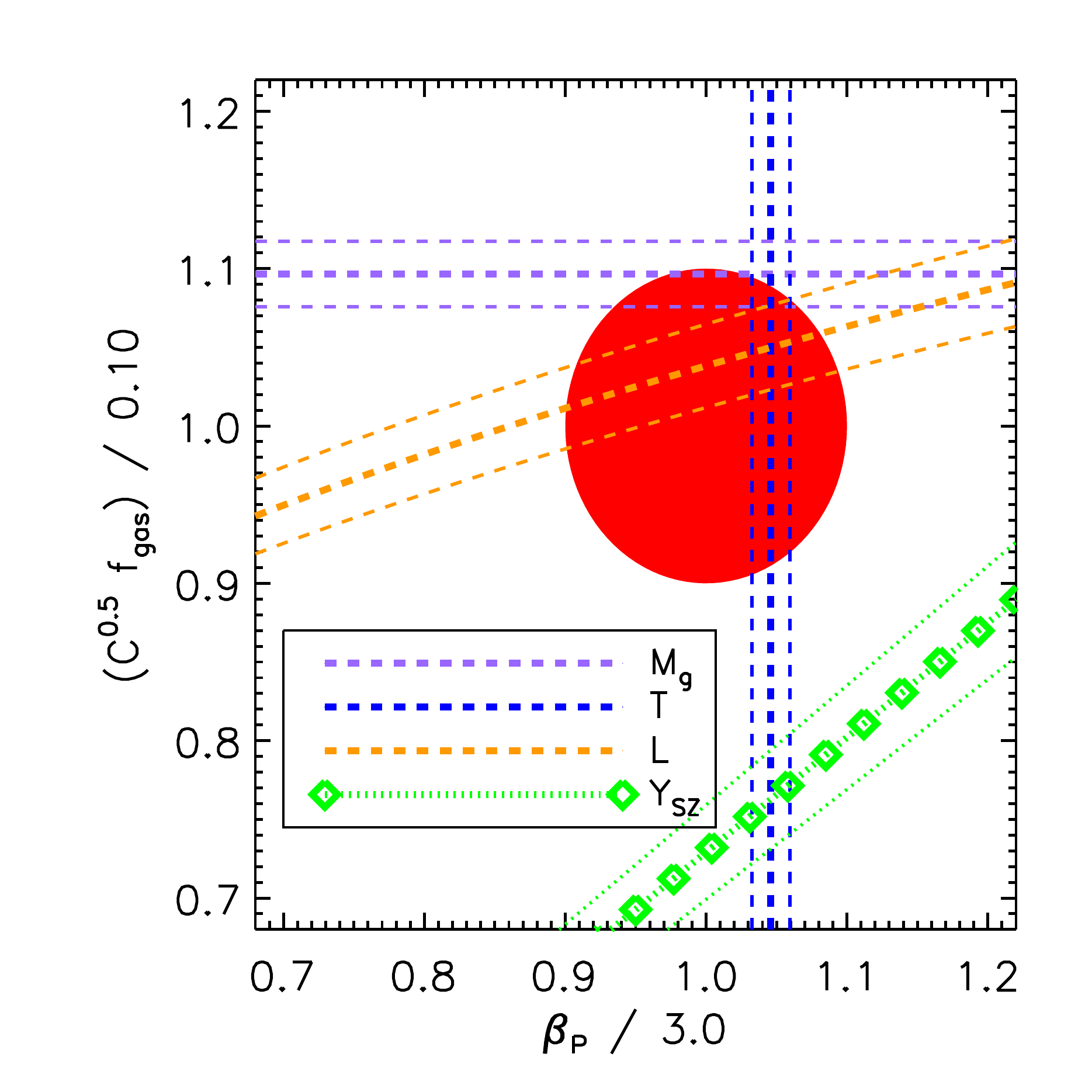}
}  \end{center}
\begin{center}
\hbox{
 \includegraphics[width=0.5\textwidth, keepaspectratio]{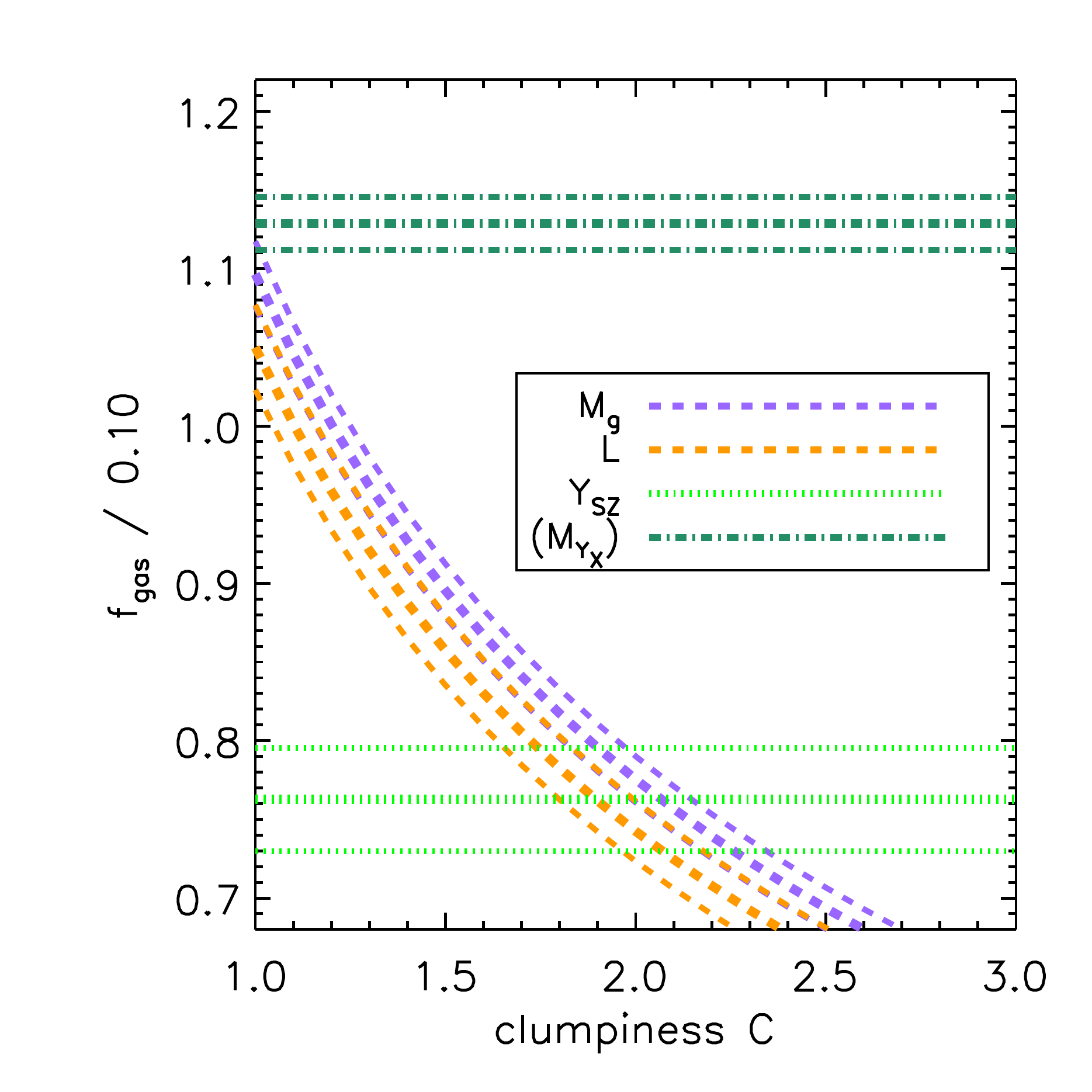}
 \includegraphics[width=0.5\textwidth, keepaspectratio]{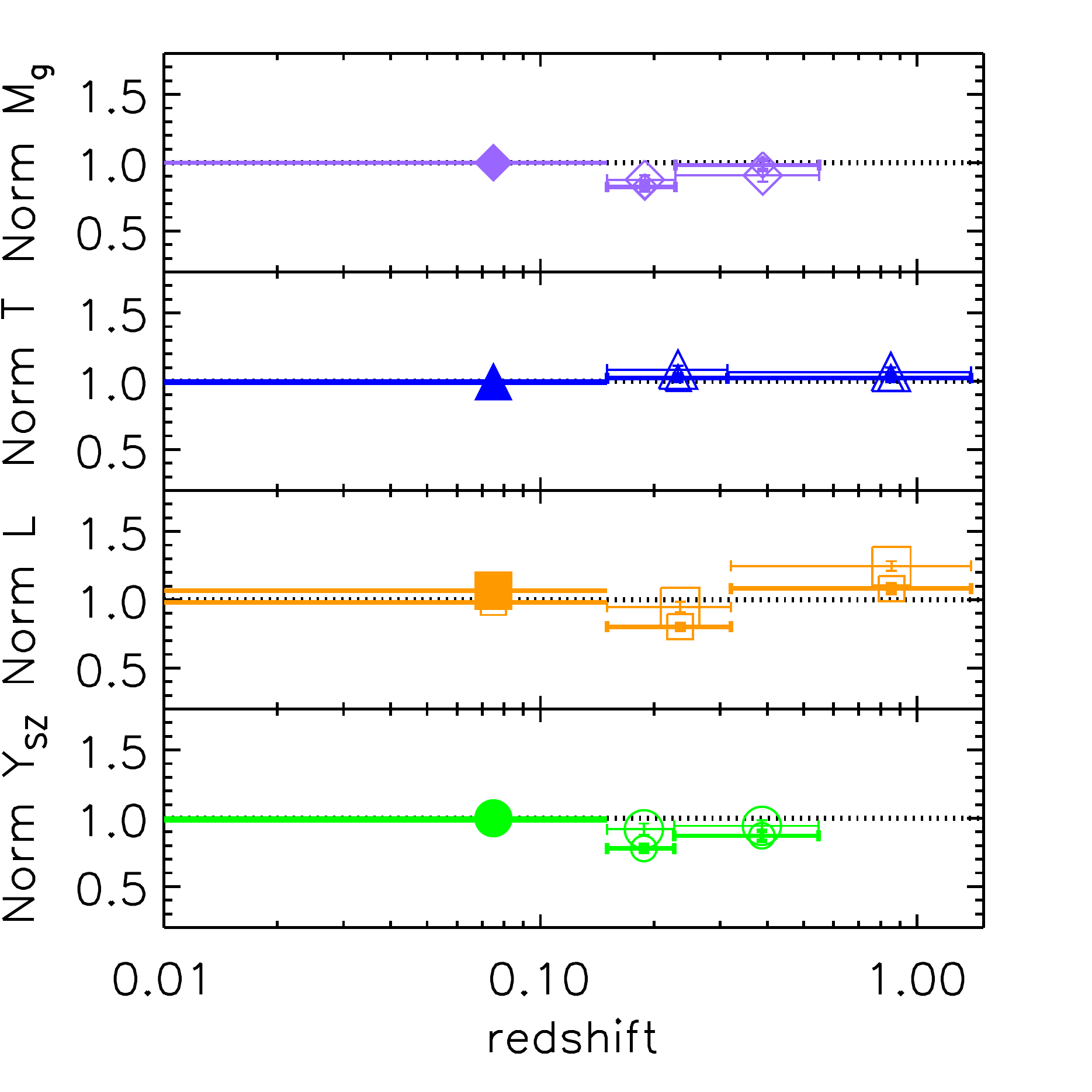}
} \end{center} \caption{From {\it (left, top)} to {\it (right, bottom)}, we show the process of the calibration of the 3 physical quantities we require to define the normalization of the scaling relations.  {\it (Left, top)} Ratios between the observed and expected normalizations of equations~\ref{eq:sl}. Fill points refer to the best-fit results measured for nearby massive systems ($M_{500}>3 \times 10^{14} M_{\odot}$ and $z<0.15$). For sake of completness, we also show (but do not use in our calculations) the ratios measured for clusters
at $z>0.15$ and divided in two redshift bins (biggest points: subsample including the half of the most massive ones) . 
{\it (Right, top)} Calibration of $C^{0.5} f_g$ and $\beta_P$ using only the nearby massive systems (fill dots in the left panel).
The red circle represents the reference values of $(\beta_P, f_g) = (3, 0.1)$ with a relative uncertainty of 10 per cent.
{\it (Left, bottom)} Combination of the constraints from the normalizations of the $M-M_g$, $M-L$ and $M-Y_{SZ}$ relations in the gas mass fraction--gas clumpiness plane. The label ``$(M_{Y_X})$'' indicates the constraint on $f_g$ obtained from the $M_{Y_X}-Y_{SZ}$ relation (see Sect.~\ref{sect:bestfit}).
{\it (Right, bottom)} As in the upper-left panel, after correcting for the best-fit values of $(\beta_P, f_g)$ in equation~\ref{eq:varm} as discussed in Sect.~\ref{sect:mass} and \ref{sect:bestfit}.
} \label{fig:norm}
\end{figure*}

\subsection{The best-fit constraints}
\label{sect:bestfit}

In an ideal case, where the samples analyzed have a well-known selection function, a direct constraint on the dependence of the scaling relations on the mass (and the redshift) could be obtained by applying the equations listed above.
In particular, from eq.~\ref{eq:sl_msingle}, one can recover 
\begin{enumerate}
\item $\beta_{P,0}$ and $m_1$ from the best-fit normalization and slope of the observed $M-T$ relation;
\item using (i), $f_{g,0}$ and $m_2$ from the best-fit normalization and slope of the observed $M-Y_{SZ}$ relation;
\item using (ii), $C_0$ and $m_3$ from the best-fit normalization and slope of the observed $M-M_g$ (or $M-L$) relation.
\end{enumerate}

However, considering that (a) our datasets have been collected from the literature (see discussion in E13) and, thus, cannot be treated as a statistically well-defined sample, and (b) a different definition of $R_{500}$ as recovered from, e.g., hydrostatic masses and $M_{Y_{X}}$ (see, for instance, the mass-dependent bias shown in Fig.~\ref{fig:planck_m}) affects the reconstructed Planck $Y_{SZ}$ signal, we decide to proceed differently.
First, we decide not to use the $M-Y_{SZ}$ relation to calibrate the gas mass fraction. This implies that we have to deal with a degeneracy between the mass dependence on $f_g$ and $C$. Therefore, we fix $m_3=0$, assuming that the gas clumpiness does not have any significant dependence on the cluster mass (see e.g. Nagai \& Lau 2011 and Roncarelli et al. 2013, where a marginal mass dependence for simulated systems appears at radii beyond $R_{200}$, but it is almost negligible at $R_{500}$).
Second, we use the whole sample of local systems (``local all" sample in Table~\ref{tab:res}) to quantify the mass dependence in eq.~\ref{eq:par_m}. 
To do that, we use equation~\ref{eq:cor} and compute the corrected values of the normalization $\mathcal{N}^c$ from the observed best-fit parameters.
The best-fit values of $m$ are quoted in Table~\ref{tab:res} and imply that
$m_1 = 0.071 \pm 0.012$ and $m_2 = 0.198 \pm 0.025$.
Third, to constrain the normalizations $\beta_{P,0}$ and $C^{0.5} f_{g,0}$, we analyze the subsamples of  the nearby ($z<0.15$), massive ($M_{500}>3 \times 10^{14} M_{\odot}$) galaxy clusters (``local massive" sample in Table~\ref{tab:res}). Doing that we minimize the effect of a mass and redshift dependence on these values, and avoid any significant Malmquist bias due to the fact that the average luminosity of selected clusters is higher than that in the parent population in a flux limited sample (e.g. Stanek et al. 2006, Pratt et al. 2009).
Proceeding in this way, we constrain $\beta_{P,0}$ and $C^{0.5} f_{g,0}$ from the best-fit normalization of the observed $M-T$ and $M-M_g$ relation, respectively, and obtain $C_0^{0.5} f_{g,0}=0.110 (\pm 0.002)$ and $\beta_{P,0}=3.14 (\pm 0.04)$, respectively, at $\Delta=500$
(see top-right panel of Fig.~\ref{fig:norm}).

For sake of completeness, we show in Fig.~\ref{fig:norm} also the ratios between the estimated normalization and the expected value obtained in two redshift bins (defined with respect to the median value in the interval $0.15-$max($z$)) and in two mass bins (build accordingly to the median value in each redshift bin).
These ratios indicate that our procedure is already capable to reproduce reasonably well the scaling relations for systems in the low-mass and/or high-redshift regime.
On the other hand, a proper treatment of these cases requires the adoption of the selection function used to define our sample. This treatment is beyond the purpose of the present work and can be avoided just considering local, and massive, objects. 

Considering now the $M-Y_{SZ}$ relation, where the normalization is independent from the clumpiness, we can break the degeneracy between $C$ and $f_g$
(see bottom-left panel of Fig.~\ref{fig:norm}) and obtain: $C_0 = 2.07 (\pm 0.02)$ and $f_{g,0} = 0.076 (\pm 0.003)$.
 
To summarize, we calibrate the new formalism in the following way:
\begin{enumerate}
\item using the ``local all" sample, we quantify the mass dependence $m_1$ and $m_2$ ($m_3$ is fixed equal to 0) using eq.~\ref{eq:sl_msingle} [see fit labelled {\it (5)} in Table~\ref{tab:res}];
\item we estimate $\beta_{P,0}$ and $C_0^{0.5} f_{g,0}$ in the ``local massive'' samples through the $M-T$ and $M-M_g$ relation, respectively, by equation~\ref{eq:fit_m}  [see fit labelled {\it (3)} in Table~\ref{tab:res}];
\item the degeneracy between $C_0$ and $f_{g,0}$ is broken with the $M-Y_{SZ}$ relation for the ``local massive'' systems.
\end{enumerate}

All the quoted errors are at $1 \sigma$ level and originate from the statistical uncertainties only.
When we take into account the uncertainties related to the cross-calibration between {\it Chandra} and {\it XMM-Newton} on the gas temperature, gas mass, gas luminosity and hydrostatic mass as discussed, e.g. in Maughan (2013) and Mahdavi et al. (2013; also private communication), systematic errors of $\pm 0.002$ and $\pm 0.02$ affect the normalization of $C^{0.5} f_g$ and $\beta_P$, respectively, whereas the error associated on the slope of the mass dependence is about $\pm 0.04$ and $\pm 0.004$, respectively.

Once we have constrained the normalisations and mass dependence of the quantities in equation~\ref{eq:par_m}, we re-estimate the ratios between the normalizations of the scaling relations and the predicted values.  As shown in  Fig.~\ref{fig:norm} (panel at the bottom-right), we obtain a match in the order of few per cent for all the set of scaling laws investigated. The fit labelled {\it (7)} in Table~\ref{tab:res} indicates the results obtained by fixing both the slope (to the self-similar expectation) and the normalization (after the calibration described above) of the scaling relations. Both the reduced $\chi^2$ and the instrinsic scatter are lower than in the scaling laws where normalizations and slopes are used as free parameters.

\subsection{Comparison with previous work}

The constraint on the value of $C^{0.5} f_g$ is perfectly consistent with the results on the gas mass fraction obtained from recent work on both X-ray observations and the most recent hydrodynamical numerical simulations.
By combining observational constraints from Vikhlinin et al. (2006), Arnaud et al. (2007) and Sun et al. (2009), Pratt et al. (2009) quote a gas mass fraction at $\Delta=500$ of  $0.113 (\pm 0.005) (M / 5 \times 10^{14} M_{\odot})^{0.21 (\pm 0.03)}$. Planelles et al. (2013), using a set of cosmological SPH hydrodynamical simulations of massive ($M_{500} > 2.8 \times 10^{14} M_{\odot}$) galaxy clusters, measure, in the redshift range 0--1, a mean gas mass fraction in the range between 0.105 (for simulations including radiative cooling, star formation and feedback from supernovae) and 0.140 (for the non-radiative set), with an average value of 0.117 (and a rms of 0.008) for the objects simulated also accounting for the effect of feedback from active galactic nuclei.
This would require $C \approx 1$, implying that the considered $Y_{SZ}$ signal is biased high by about $(0.110/0.076) \sim 45$ per cent at given mass.
This amount is difficult to explain with some selection effect, also considering that twenty-two (out of 27) of the systems included in the ``local massive" sample have a signal-to-noise ratio related to the SZ detection in correspondence of the X-ray position larger than 7 (all the local, massive objects have a signal-to-noise ratio in the range 5.7--26.5, with a median value of 10.3), making them less prone to any Malmquist-like bias propagated through the sample selection (see, e.g., discussion in Sect.~7.5.2 of Planck collaboration 2014). 
On the other hand, if we replace the hydrostatic masses with the values $M_{Y_{X}}$ estimated through the $Y_X$ parameter (see a discussion on the comparison between them in Sect.~\ref{sect:samples} and Fig.~\ref{fig:planck_m}) and fit equation~\ref{eq:fit_m}, we measure $\mathcal{N}_{\rm obs} = 1.670 (\pm 0.008)$ that implies a gas mass fraction of $0.113 (\pm 0.002)$ and, combined with the result on the $M-M_g$ relation, a gas clumpiness slightly lower than the physically motivated lower bound of 1 ($C \sim 0.95$). 
We conclude that, for the available dataset, some tension between hydrostatic $M$ and $M_{Y_{X}}$ is present (see, for instance, the mass-dependent bias shown in Fig.~\ref{fig:planck_m}) that does not permit to break univocally the degeneracy between $f_g$ and $C$.
We recognize also that more work on this topic, with a more extended and detailed comparison between hydrostatic masses and integrated Compton parameters, is needed, but beyond the purpose of the present study.

\begin{figure}
 \includegraphics[width=0.5\textwidth, keepaspectratio]{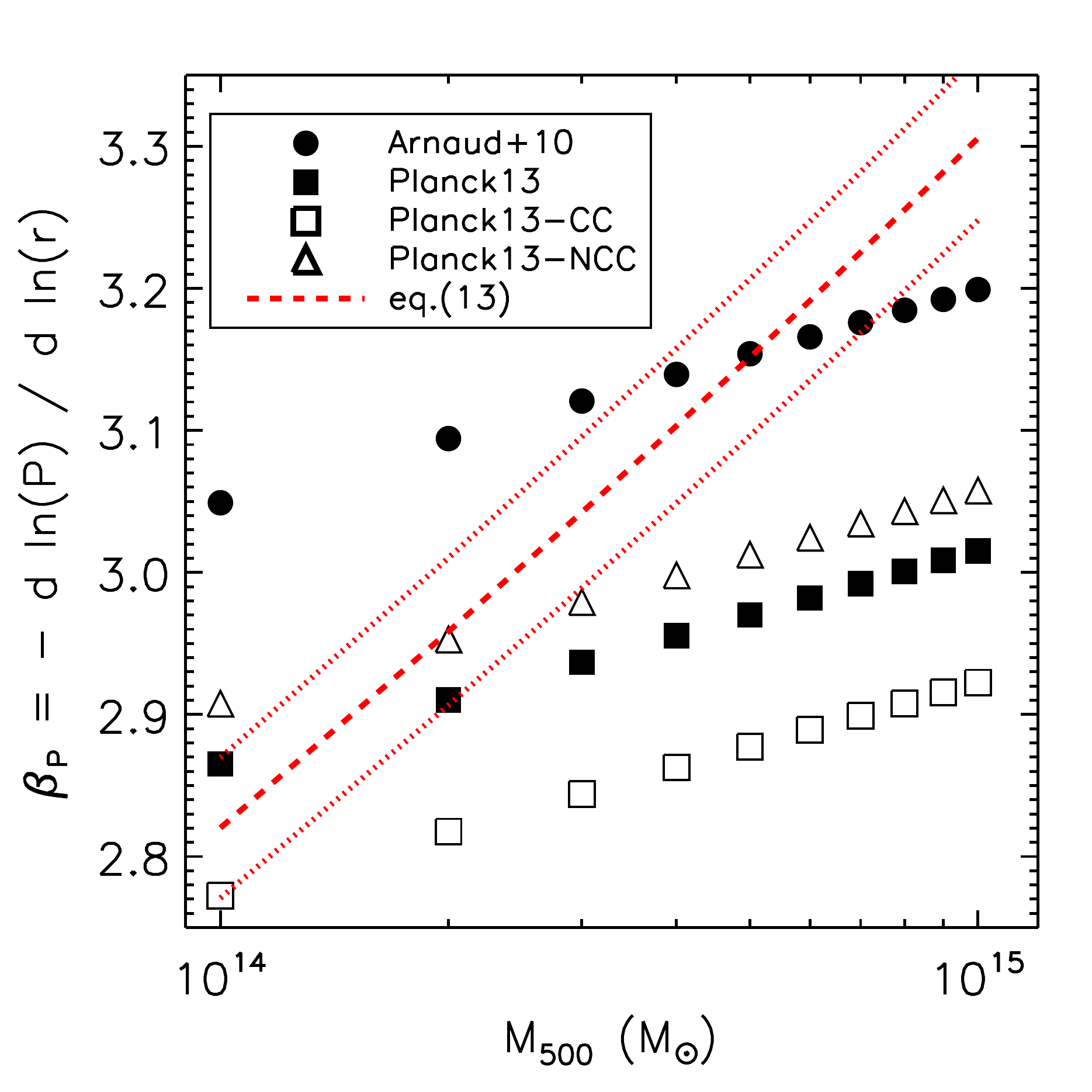} 
 \caption{Constraints on the logarithmic slope of the pressure profile as function of $M_{500}$.
 The dotted lines show the $1 \sigma$ uncertainty associated to the best fit result (dashed line; see eq.~\ref{eq:varm}). 
The points refer to the best-fit models adopted in Planck Intermediate Results (2013) and Arnaud et al. (2010) as labelled.
}
\label{fig:bp}
\end{figure}

The mass dependence of the gas pressure profile (see results for the $M-T$ relation for the ``local all" sample in Tab.~\ref{tab:res}) is not in contrast with the present observational constraints (e.g. Arnaud et al. 2010, Planck  collaboration 2013, Sun et al. 2011). In Fig.~\ref{fig:bp}, we show our best-fit constraints compared to the predictions from the best-fit values of the universal model presented in Arnaud et al. (2010) and in Planck Intermediate Results (2013). This universal model is obtained by combining observational data based on \xmm\ observations in the radial range 0.03--1 $R_{500}$ with hydro-simulations results out to 4  $R_{500}$ and using a generalized NFW functional form (originally proposed by Nagai et al. 2007) to fit the combined re-scaled profile.
Our result on the logarithmic slope of the pressure profile at $R_{500}$, $\beta_P$, shows a steeper mass dependence, with values that lie
between $2.8$ at $\sim 10^{14} M_{\odot}$, preferred also from the Planck collaboration best-fit parameters, and $3.3$ at  $\sim 10^{15} M_{\odot}$, more in agreement with the Arnaud et al. profile. 

\section{Summary and discussion}

In the present work, we estimate the predicted values of the normalization and slope of the scaling relations holding between the
hydrostatic mass and (i) the gas mass, (ii) the gas temperature, (iii) the X-ray bolometric luminosity, (iv) the integrated Compton parameter.
We show in details how these normalizations depend upon the gas density clumpiness $C$, the gas mass fraction $f_g$ and the logarithmic slope of the thermal 
pressure profile $\beta_P$. 
We argue that the deviations of the observed slopes from the self-similar expectations can be fully explained with a mass dependence 
of the gas mass fraction and the logarithmic slope of the thermal pressure profile. 

Relying on the availability of large database of measured hydrostatic masses and observables in X-ray and millimeter wave bands,
we constrain at high significance the normalization and mass dependence of the gas mass fraction and the logarithmic slope of the thermal pressure profile, putting also some limits on the level of gas clumpiness requested to accommodate in a self-consistent scenario all the set of the scaling relations.
We conclude that 
\begin{enumerate}
\item the 3 astrophysical quantities (i.e. gas clumpiness, gas mass fraction and slope of the pressure profile) advocated to explain consistently the predicted $M - \left\{ M_g, T, L, Y_{SZ} \right\}$ relations are sufficient to define the observed normalization and slope of these scaling laws;
\item using nearby ($z<0.15$), massive ($M_{500}>3 \times 10^{14} M_{\odot}$) galaxy clusters, the $M-M_g$ requires $(C^{0.5} f_g) = 0.110 (\pm 0.002)$. Using the further constraint obtained from the clumpiness-free normalization of the $M-Y_{SZ}$ relation, we obtain that, within $R_{500}$,
the gas clumpiness  is $2.07 (\pm 0.02)$ and the gas mass fraction is $0.076 (\pm 0.003)$ (see Fig.~\ref{fig:norm});
\item we note, however, that being the constraint on $(C^{0.5} f_g)$ well in agreement with results from, e.g., recent hydrodynamical simulations on the cluster gas mass fraction at $\Delta=500$ (e.g. Planelles et al. 2013), it would suggest that $C \approx 1$ and that the considered $Y_{SZ}$ signal is biased high by $(0.110/0.076) \sim 45$ per cent at given mass;
\item considering that most of the galaxy clusters  included in the ``local massive" sample have a signal-to-noise ratio related to the SZ detection in correspondence of the X-ray position larger than 7, we exclude any significant Malmquist-like bias affecting the analyzed sample;
\item on the other hand, if we replace the hydrostatic masses with the values obtained from the Planck collaboration through the $Y_X$ parameter and carry on the same analysis, we obtain indeed that $C \sim 1$. However, this result highlights a tension between the measurements of $M$ and $M_{Y_{X}}$ for the same objects, with the  hydrostatic estimates that over (under) predict the high (low) values of $M_{Y_X}$ by about 30 per cent (see Fig.~\ref{fig:planck_m});
\item using the same sample of local and massive galaxy clusters and the $M-T$ relation, we constrain $\beta_P=  -d \ln P/d \ln r  = 3.14 (\pm 0.04)$;
\item we quantify the dependence upon the mass of the 2 adopted quantities (the clumpiness is assumed be independent from the mass, i.e. $m_3=0$) through the best-fit parameters of equation~\ref{eq:sl_msingle}, and obtain: $f_g \sim M^{0.20\pm0.02}$ and $\beta_P \sim M^{0.07 \pm 0.01}$; 
while the former is in good agreement both with other observational results and profiles predicted from hydrodynamical simulations, the latter one shows agreement with the Planck collaboration (2013) best-fit parameters at lower ($\sim 10^{14} M_{\odot}$) masses and with the Arnaud et al. (2010) profile at higher ($\sim 10^{15} M_{\odot}$) masses, requiring a steeper mass dependence;
\item by adjusting for the mass dependence of $f_g$ and $\beta_P$, we demonstrate [see results labelled with ``eq.~\ref{eq:fit_m}" and fit {\it (7)} in Table~\ref{tab:res}] that the scaling relations with a slope fixed to the expected value in the self-similar scenario provide best-fit results with a reduced $\chi^2$ and an intrinsic scatter comparable to the results obtained leaving the slope free to vary.
\end{enumerate}

Therefore, we conclude that the scaling relations based on X-ray/SZ quantities have a simple and predictable behavior that can be fully described at $\Delta=500$ by the equations~\ref{eq:sl} and \ref{eq:par_m} (or their formal extension in equation~\ref{eq:sl_m}), where 
\begin{align}
C^{0.5} & f_g = 0.110 (\pm 0.002) \left( \frac{E_z M}{5 \times 10^{14} M_{\odot}} \right)^{0.198 (\pm 0.025)} \nonumber \\
\beta_P = & -\frac{d \ln P}{d \ln r} =  3.14 (\pm 0.04)  \left( \frac{E_z M}{5 \times 10^{14} M_{\odot}} \right)^{0.071 (\pm 0.012)}.
\label{eq:varm}
\end{align}
The quoted uncertainties are statistical only and are the products of the propagation of the relative error available to the estimates of the hydrostatic masses, gas masses, temperature and luminosity and the size of the cluster sample analzyed. When the uncertainties related to the cross-calibration between {\it Chandra} and {\it XMM-Newton} on the gas temperature and hydrostatic mass is taken into account as discussed, e.g. in Maughan (2013) and Mahdavi et al. (2013), systematics errors in the order of (i) $\pm 0.002$ and $\pm 0.04$ and (ii) $\pm 0.02$ and $\pm 0.004$ affect the normalization and the slope of the mass dependence of $C^{0.5} f_g$ and $\beta_P$, respectively.

Inserting these values into equation~\ref{eq:fit_m}, the gravitating mass can be recovered with, for instance, a lower intrinsic scatter associated to it than the one measured by using the standard relations with normalization and slope free to vary.

For the set of the four relations here investigated, these results provide a significant simplification in terms of number of free parameters to be constrained: routinely, a slope and a normalization have to be estimated (for a total of 8 free parameters), whereas in our new framework, one needs only to limit the normalization of $C$, $f_g$ and $\beta_P$ and the mass dependence of the latter two, for a total of 5 free parameters.
This evidence can also be formalized by the estimates of the Akaike Information Criterion (AIC, Akaike 1974), or equivalent Information Criteria (see e.g. Liddle 2007). All our models that adopt the self-similar scaling laws with the mass-dependent physical quantities perform significantly better (from a statistical point of view) than the power-law fits where normalization and slope are left free to vary (apart from the $M-T$ relation of the sample ``local massive'', where AICs are comparable) with an evidence ratio $e^{0.5 \Delta}$, with $\Delta$ being the difference between the AIC estimated for ``free parameters" model and the one for the modified scaling relations, larger than 600 [compare, e.g.,  fits labelled {\it (3)} and {\it (6)} with the ones labelled {\it (2)} and {\it (5)} in Table~\ref{tab:res}]. 
When the 4 scaling relations are considered together, we obtain a cumulative $\chi^2$ of 1743 and 379 with 228 data points for the ``local all'' sample (222, 173 and 94, respectively, for the ``local massive'' one) for the set of the scaling laws with 8 (all the normalizations and slopes) and 5 free parameters, respectively, implying a ``decisive'' evidence (according to the Jeffreys' scale in Kass \& Raftery 1995) in favor of our alternative scenario.

We also note that the formalism described in Sect.~\ref{sect:mass} (e.g. equation~\ref{eq:sl_m}) is ready to accomodate the redshift evolution of the scaling relations through the assumed expressions in eq.~\ref{eq:par_m}. As we present in Fig.~\ref{fig:norm}, preliminary plots that do not consider any selection function show encouraging agreements between the observed distributions and the expected ones. More dedicated work to characterize properly the studied samples both as function of mass and redshift (for instance, to measure the relative weight of low-mass and high-redshift systems in the fit of the scaling relations) is however needed.

The result of this study opens a very-promising prospective to have a full set of inter-correlated and internally-consistent scaling relations that rely on the ones predicted from the self-similar scenario with an extension depending on well-identified astrophysical properties that can be investigated independently (like, e.g., the mass dependence of the thermal pressure profile or of the gas mass fraction).

\section*{ACKNOWLEDGEMENTS} 
We thank the anonymous referee for helpful comments that improved the presentation of the work.  
We thank Mauro Sereno,  Marco De Petris, Dominique Eckert for discussion and comments on the manuscript.
We acknowledge the financial contribution from contracts ASI-INAF I/009/10/0 and PRIN-INAF 2012.

\appendix
\section{Numerical estimates of the normalization}
\label{sect:nor}
For sake of completeness, we provide here the details on how the numbers of equation~\ref{eq:sl} are obtained.
Let us define 
\begin{align}
k_0 & = \frac{4}{3} \pi \Delta \rho_{c, 0} = \frac{\Delta H_0}{2 G} =1.928 \times 10^{-26} \frac{\Delta}{500} {\rm g} \, {\rm s} \, {\rm cm}^{-3} \nonumber \\
k_1 & = \mu m_{\rm amu} G k_0^{1/3} = 1.820 \times 10^{-40} {\rm g}^{1/3} {\rm s}^{-5/3} {\rm cm}^2 .
\end{align}
Then, the normalization for the $M-M_g$, $M-T$, $M-L$ and $M-Y_{SZ}$ relations can be estimated as:
\begin{align}
n_{MM_g} & = (C_0^{0.5} f_{g,0})^{-1} \frac{M_{g,0}}{M_0} \nonumber \\
n_{MT} & =  \left( \frac{f_T}{k_1} \right)^{3/2}  \beta_{P,0}^{3/2} \, \frac{T_0^{3/2}}{M_0}  \nonumber \\
n_{ML} & = \left( \frac{4/3 \pi \, \mu_e^2 \, m_{\rm amu}^2 \, f_T^{0.5}}{f_L \, c_{f,0} \,  k_0 \, k_1^{0.5}} \right)^{3/4}  \beta_{P,0}^{3/8} (C_0^{0.5} f_{g,0})^{-3/2}
  \, \frac{L_0^{3/4}}{M_0} \nonumber \\
n_{MY} & = \left( \frac{m_e \, c^2 \, \mu_e \, m_{\rm amu} \, f_T}{k_1 \sigma_T} \right)^{3/5} \beta_{P,0}^{3/5} f_{g,0}^{-3/5} \, \frac{Y_0^{3/5}}{M_0},
\end{align}
where $M_0$, $M_{g,0}$, $T_0$, $L_0$ and $Y_0$ are the pivot values in c.g.s unit and are equal to $5 \times 10^{14} M_{\odot}$, $5 \times 10^{13} M_{\odot}$
$5$ keV, $5 \times 10^{44}$ erg s$^{-1}$ and $10^{-4}$ Mpc$^2$, respectively, in the present work.

\section{Energy band dependence of the $M-L$ relation}
\label{sect:ml_eband}
The gas luminosity considered in our analysis is the X-ray bolometric one, i.e. it has been evaluated in the energy band 0.01--100 keV.
We indicate here how the $M-L$ relation is modified once the luminosity is estimated in different energy bands.
In these cases, the cooling function $c_f$ will not show a dependence upon the temperature to the power of $1/2$. 
By approximating the cooling function as a power-law of the temperature, we can write $c_f = c_{f,0} \times T^{\tau}$ and
\begin{align}
\frac{F_z M}{5 \times 10^{14} M_{\odot}}  & = n_{MLe} \left( \frac{\beta_P}{3} \right)^{\tau/ (1+2 \tau/3)} \left( \frac{C^{0.5} \, f_g}{0.1} \right)^{-2 /(1+2 \tau/3)}  \nonumber \\
 \hspace*{2cm} & \left( \frac{F_z^{-1} L}{5\times10^{44} {\rm erg/s}} \right)^{1 / (1+2 \tau/3)},  \nonumber \\
n_{MLe} & = \left( \frac{4/3 \pi \, \mu_e^2 \, m_{\rm amu}^2 \, f_T^{\tau}}{f_L \, c_{f,0} \,  k_0 \, k_1^{\tau}} \right)^{1 / (1+2 \tau/3)}.
\end{align}

We quote here $c_{f,0}$, $\tau$ and the modified $M-L$ relation for the most commonly used energy bands: 
\begin{enumerate}
\item {\bf (0.1--2.4 keV)} formally, the best-fit values with a power-law of the cooling function in the range 2--12 keV are $c_f = 1.12   \times 10^{-23} c_{pe} T_{\rm keV}^{-0.11}$ erg s$^{-1}$ cm$^{3}$. Adopting an exponent $\tau=0$, $c_{f,0} = 0.91 \times 10^{-23} c_{pe}$ and the $M-L$ relation can be written as
\begin{equation}
\frac{F_z M}{5 \times 10^{14} M_{\odot}}  = 2.110  \left( \frac{C^{0.5} \, f_g}{0.1} \right)^{-2}  
\left( \frac{F_z^{-1} L}{5\times10^{44} {\rm erg/s}} \right) 
\end{equation}

\item  {\bf (0.5--2 keV)}  As above, $c_f = 0.68 \times 10^{-23} c_{pe} T_{\rm keV}^{-0.10}$ erg s$^{-1}$ cm$^{3}$. 
With a null dependence upon the temperature, $c_f = c_{f,0} = 0.56 \times 10^{-23} c_{pe}$ and the $M-L$ relation can be written as
\begin{equation}
\frac{F_z M}{5 \times 10^{14} M_{\odot}}  = 3.400  \left( \frac{C^{0.5} \, f_g}{0.1} \right)^{-2}  
\left( \frac{F_z^{-1} L}{5\times10^{44} {\rm erg/s}} \right) 
\end{equation}

\item  {\bf (2--10 keV)} In this case, $c_f = 0.38 \times 10^{-23} c_{pe} T_{\rm keV}^{0.5}$ erg s$^{-1}$ cm$^{3}$. Then, 
\begin{align}
\frac{F_z M}{5 \times 10^{14} M_{\odot}}  = & 1.749 \left( \frac{\beta_P}{3} \right)^{3/8} \left( \frac{C^{0.5} \, f_g}{0.1} \right)^{-3/2}  \nonumber \\
\hspace*{2cm} & \left( \frac{F_z^{-1} L}{5\times10^{44} {\rm erg/s}} \right)^{3/4}.
\end{align}
\end{enumerate}

\section{Plots of the investigated scaling relations}

We collect here the plots, with the best-fit lines and the corresponding residuals $\chi_i$ of equation~\ref{eq:chi2}, of the samples described in Table~\ref{tab:res}.
The normalizations of these best-fit scaling relations are shown in Fig.~\ref{fig:norm}.

\begin{figure*}
\vspace*{-3cm} \hbox{
 \includegraphics[width=0.5\textwidth, keepaspectratio]{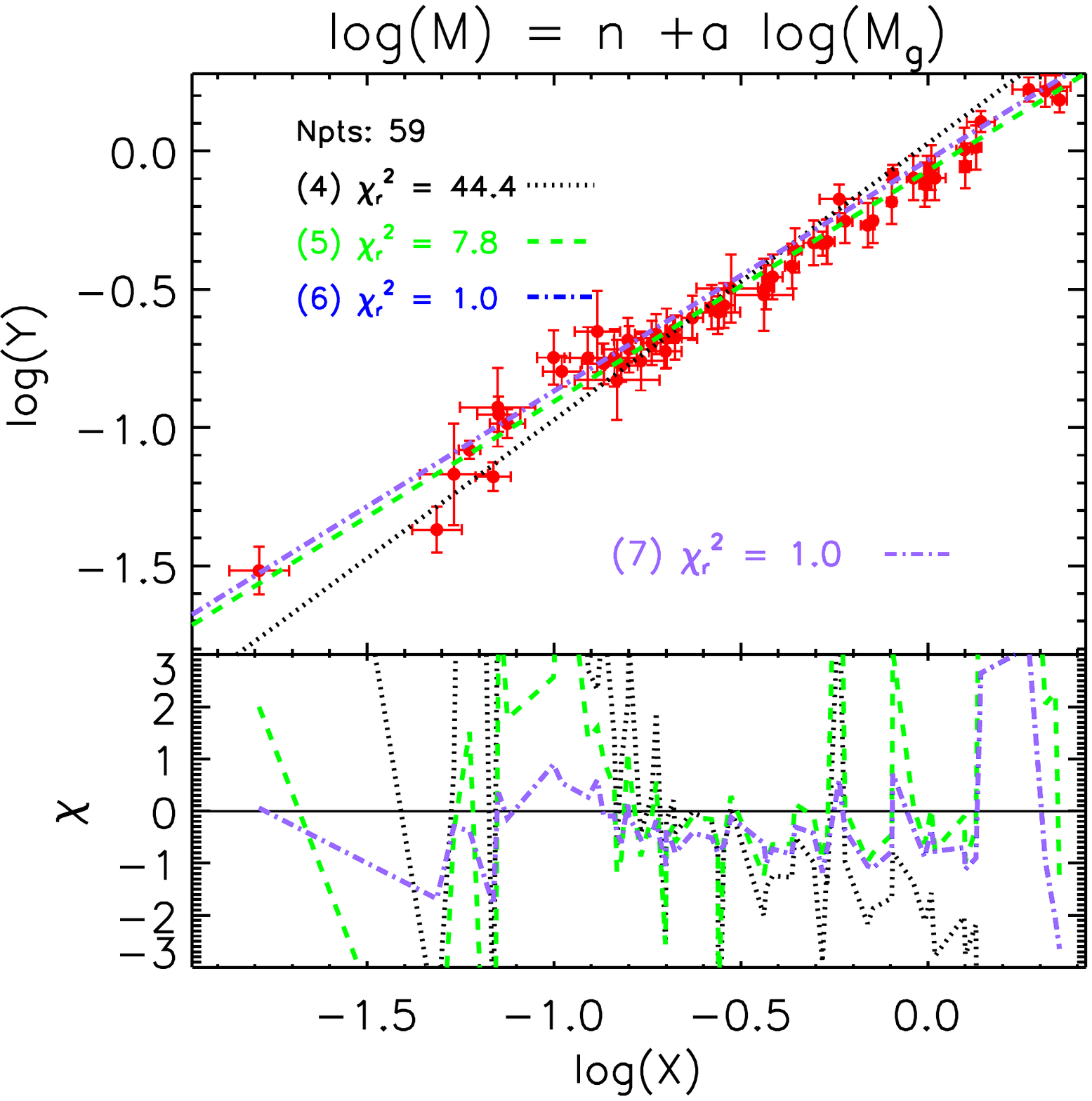}
  \includegraphics[width=0.5\textwidth, keepaspectratio]{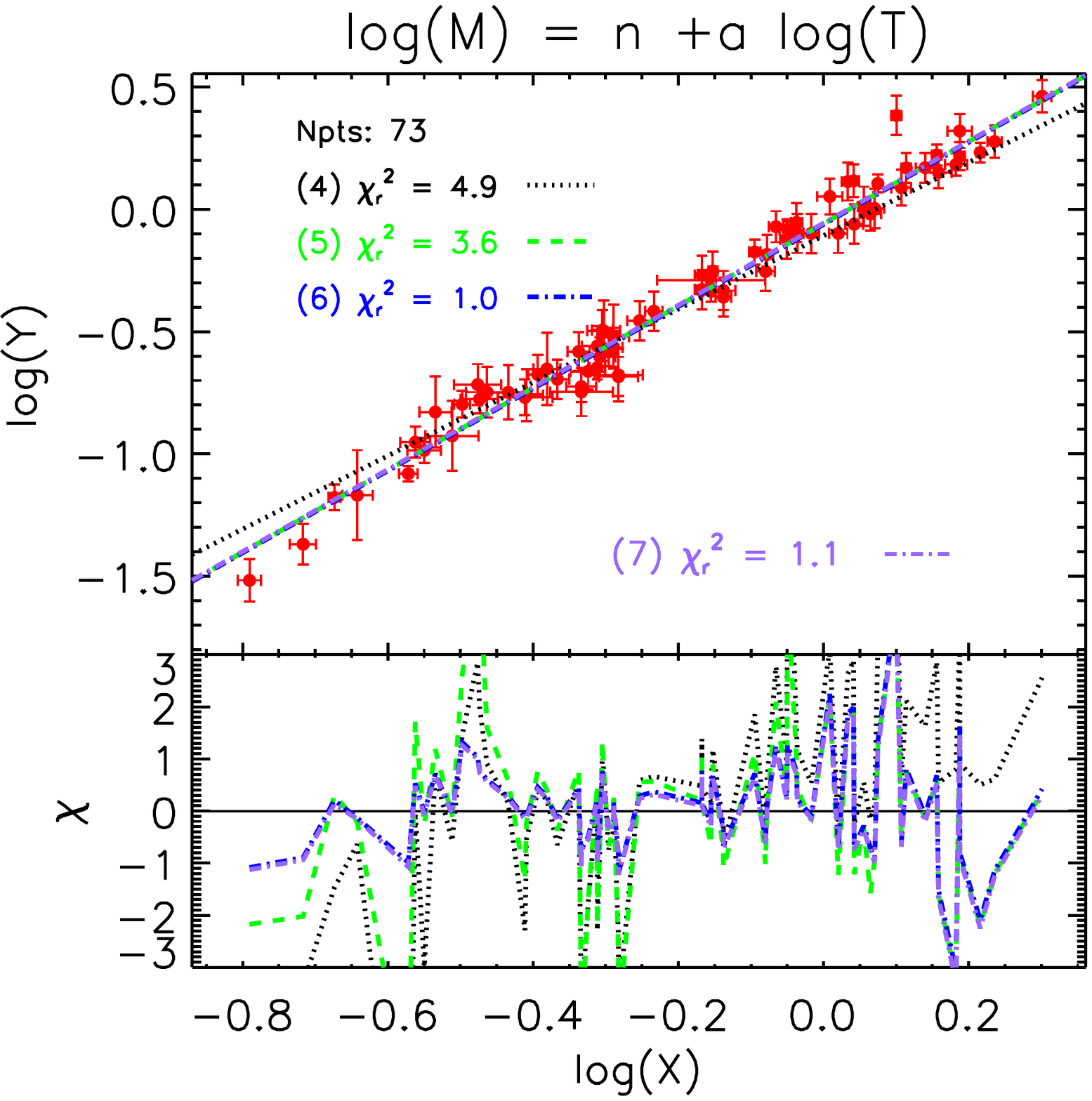}
  } \vspace*{-3cm} \hbox{
    \includegraphics[width=0.5\textwidth, keepaspectratio]{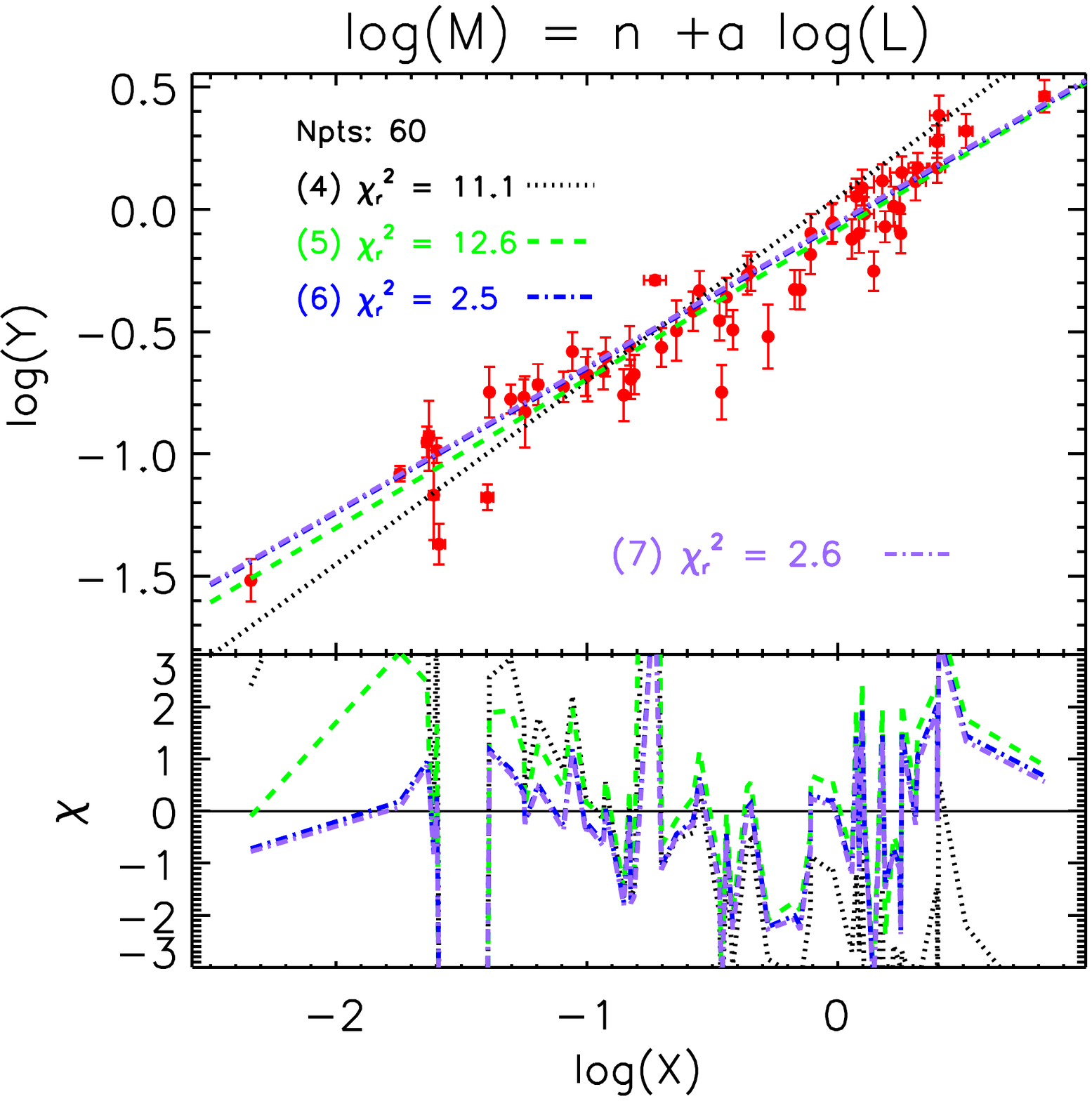}
  \includegraphics[width=0.5\textwidth, keepaspectratio]{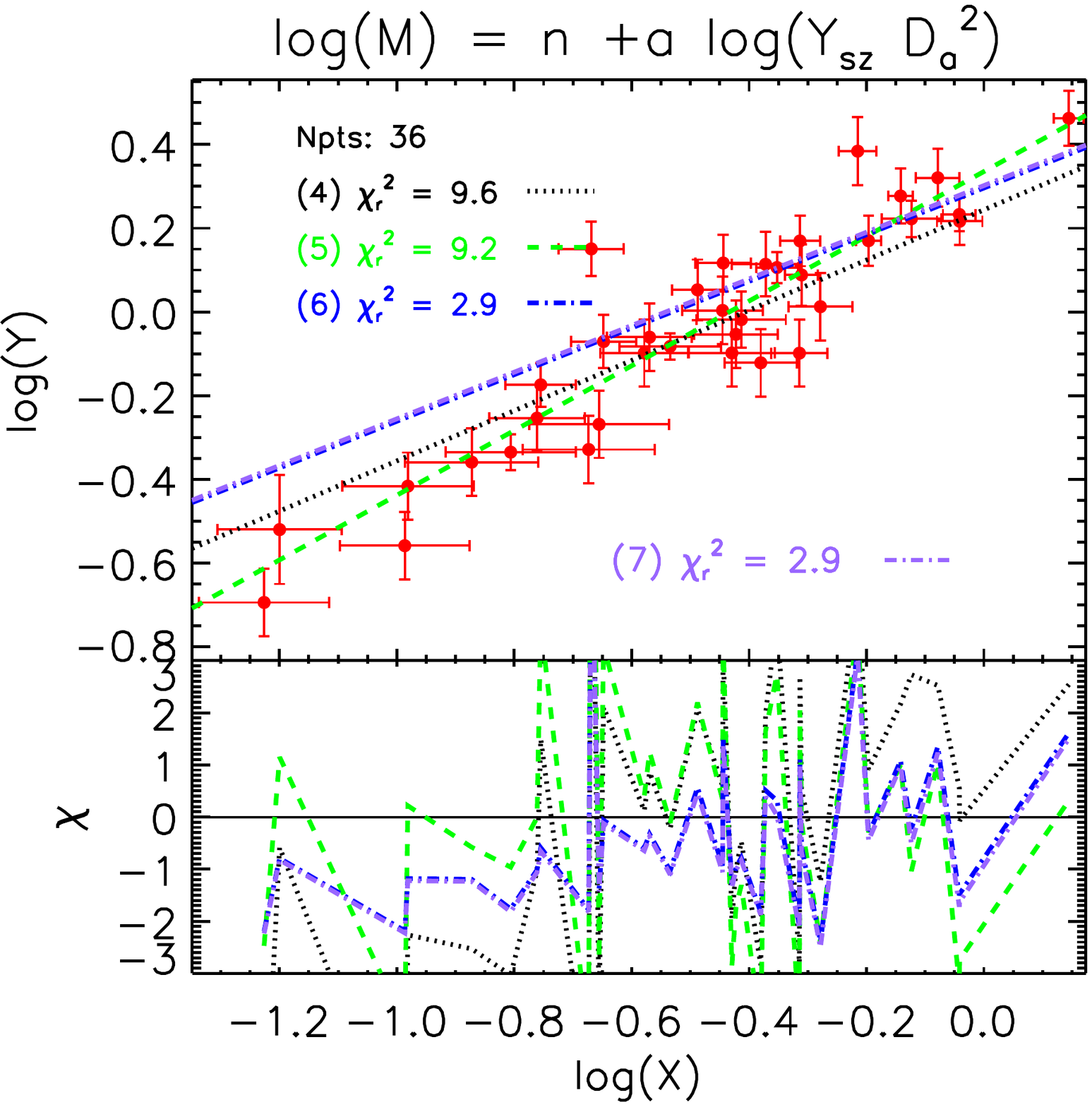}
}  \caption{These plots show the samples and the best-fit results, with the associated residuals $\chi$, described from lines {\it (4), (5), (6), (7)} in Table~\ref{tab:res}.
Note that the best-fit lines labelled {\it (6)} and {\it (7)} have been corrected by the factor $(1 -\theta \; m_1 + \phi \; m_2 +\phi \; m_3/2)$ in equation~\ref{eq:fit_m} for the sake of representation. The sum of the squared residuals $\chi$ provides the quoted total $\chi^2$.
} \label{fig:fit}
\end{figure*}

 \end{document}